\documentclass{cambridge6A}
\usepackage{graphicx}

\usepackage{amssymb}

\newcommand{\labell}[1]{\label{#1}}
\def\({\left(} \def\){\right)}
\def\[{\left[} \def\]{\right]}
\def\al{\alpha} 

\newcommand{\non}{\nonumber \\}

\newcommand{\be}{\begin{equation}}
\newcommand{\ee}{\end{equation}}
\newcommand{\bea}{\begin{eqnarray}}
\newcommand{\eea}{\end{eqnarray}}
\newcommand{\ba}{\begin{eqnarray}}
\newcommand{\ea}{\end{eqnarray}}

\newcommand{\beq}{\begin{equation}}
\newcommand{\eeq}{\end{equation}}
\newcommand{\beqa}{\begin{eqnarray}}
\newcommand{\eeqa}{\end{eqnarray}}
\newcommand{\beqar}{\begin{eqnarray*}}
\newcommand{\eeqar}{\end{eqnarray*}}

\newcommand{\reef}[1]{(\ref{#1})}

\newcommand{\eg}{{\it e.g.,}\ }
\newcommand{\ie}{{\it i.e.,}\ }
\newcommand{\comment}[1]{{\bf [[[#1]]]}}

\newcommand{\mt}[1]{\textrm{\tiny #1}}

\newcommand{\ga}{\gamma}

\newcommand{\cA}{{\cal A}}

\newcommand\dd{$d$}
\newcommand\rh{r_\mt{H}}

\begin{document}

\chapterauthor{Robert C. Myers}

\chapter{Myers-Perry black holes}

\contributor{Robert C. Myers \affiliation{Perimeter Institute}}

In this chapter, we will continue the exploration of black holes in
higher dimensions with an examination of asymptotically flat black
holes with spherical horizons, \ie in $d$ spacetime dimensions, the
topology of the horizon and of spatial infinity is an $S^{d-2}$. In
particular, we will focus on a family of vacuum solutions describing
spinning black holes, known as Myers-Perry (MP) metrics. In many
respects, these solutions admit the same remarkable properties as the
standard Kerr black hole in four dimensions. However, studying these
solutions also begins to provide some insight into the new and unusual
features of event horizons in higher dimensions.

These metrics were discovered in 1985 as a part of my thesis work as a
Ph.D. student at Princeton \cite{merry}. My supervisor, Malcolm Perry,
and I had been lead to study black holes in higher dimensions, in part,
by the renewed excitement in superstring theory which had so
dramatically emerged in the previous year. We anticipated that
examining black holes in $d>4$ dimensions would be important in
obtaining a full understanding of these theories. I should add that
amongst the subsequent developments, this family of spinning black hole
metrics was further generalized to include a cosmological constant, as
well as NUT parameters.\footnote{There is more than one such parameter
in higher dimensions.} While I will not have space to discuss these
extensions, the interested reader may find a description of the
generalized solutions in ref.~\cite{general}.

\section{Static Black Holes}

Before considering spinning black holes, we should mention that the
Schwarzschild solution is easily generalized to $d\ge4$ dimensions as
 \be
ds^2=-\left(1-\frac{\mu}{r^{d-3}}\right)\,dt^2+
\left(1-\frac{\mu}{r^{d-3}}\right)^{-1} dr^2+r^2\,d\Omega^2_{d-2}
 \labell{sch}
 \ee
where $d\Omega^2_{d-2}$ denotes the line element on the unit
($d$--2)-sphere. While this vacuum solution of the $d$-dimensional
Einstein equations was first found by Tangherlini in the early 1960's
\cite{tang}, it is still traditionally referred to as a Schwarzschild
black hole. In part, this nomenclature probably arose because for any
value of $d\,>\, 4$, the features of this spacetime \reef{sch} are
essentially unchanged from its four-dimensional predecessor.

In particular, the constant $\mu$ emerges as an integration constant in
solving the Einstein equations. In Appendix A, we derive expressions
for the mass and angular momentum in a $d$-dimensional spacetime by
examining the asymptotic structure of the metric. There one finds that
$\mu$ fixes the mass of the black hole \reef{sch} --- see
eq.~\reef{pert1}
--- with
 \be
M= {(d-2)\,\Omega_{d-2}\over16\pi G}\ \mu
 \labell{massch}
 \ee
where $\Omega_{d-2}$ is the area of a unit (\dd--2)-sphere, \ie
 \be
\Omega_{d-2}={2\,\pi^{\frac{d-1}2}\over\Gamma\!\left(\frac{d-1}{2}
\right)}\, .
 \labell{solid}
 \ee
As long as $\mu\!>\!0$, the surface $r^{d-3}$=$\mu$ is an event
horizon. It is a straightforward exercise to generalize the
discussion presented in Chapter 1 
in constructing good coordinates across this surface and finding the
maximal analytic extension of the geometry. The corresponding Penrose
diagram then takes precisely the same form as given in Figure 1.1 of
Chapter 1 
where each point now represents a (\dd--2)-sphere.\footnote{Of course,
the past and future horizons should now be labeled as $r^{d-3}$=$\mu$.}
Notably, there is a future (past) curvature singularity at $r=0$ in
region II (III), where $R_{\mu\nu\rho\sigma} R^{\mu\nu\rho\sigma}
\propto \mu^2/r^{2(d-1)}$ as $r\to 0$. Of course, if $\mu\!<\!0$ the
space-time has a naked time-like singularity at $r$=0 and the
corresponding Penrose diagram matches that
given in Figure 1.4 of Chapter 1. 

Another simple exercise is to extend Birkhoff's theorem to higher
dimensions. That is, one can solve Einstein's vacuum equations in any
$d\ge4$ with the assumption that the geometry is asymptotically flat
and spherically symmetric, \ie the solution has an $SO(d-1)$ isometry,
but without assuming that the spacetime is static. The
Schwarzschild-Tangherlini metric \reef{sch} remains the most general
solution and so any spherically symmetric solution of $R_{\mu\nu}=0$
must also be static. It is also possible to prove a uniqueness theorem
indicating that this metric \reef{sch} is the only solution of the
vacuum Einstein equations in higher dimensions if one assumes that the
geometry is asymptotically flat and static \cite{static}. Hence all
such static solutions are spherically symmetric and completely
determined by their mass $M$.

The generalization of the four-dimensional Reissner-Nordstr\"om metric
to solutions describing static charged black holes in higher dimensions
is also straightforward. Again, the features of these solutions of the
Einstein-Maxwell equations in $d>4$ are essentially unchanged from
those described for four dimensions in Chapter 1. 
Here it is interesting to extend the Majumdar-Papapetrou solutions,
describing multiple extremally charged black holes in static
equilibrium, to higher dimensions. With these solutions, one can
construct periodic arrays of such black holes which can then be
compactified using the Kaluza-Klein ansatz \cite{majp}, discussed in
Chapter 4. The resulting solutions provide simple analytic metrics
describing black holes localized in Kaluza-Klein dimensions.

\section{Spinning Black Holes} \labell{spinbh}

Before writing the metric for a spinning black hole, it is useful to
first orient the discussion by writing the metric for flat space in
higher dimensions. To begin, consider the case $d=2n+1$ (with $n\ge2$),
in which case the flat space metric can be written as
 \bea
ds^2&=& -dt^2 +\sum_{i=1}^n \left( dx_i^{\,2}+dy_i^{\,2}\right)
 \non
&=& -dt^2 + dr^2 + r^2\,\sum_{i=1}^n \left(
d\mu_i^{\,2}+\mu_i^2\,d\phi_i^{\,2}\right)\,.
 \labell{flat}
 \eea
In the first line, we have paired all of the spatial coordinates as
Cartesian coordinates ($x_i,y_i$) in $n$ orthogonal planes. In the
second line, we have introduced polar coordinates which can be
expressed with:
 \be
 x_i=r\, \mu_i\,\cos\phi_i\,,\qquad y_i= r\, \mu_i\,\sin\phi_i\,.
 \labell{polar}
 \ee
Implicitly, we are defining $r^2=\sum_{i=1}^n\left( x_i^{\,2} +
y_i^{\,2} \right)$ and so the direction cosines $\mu_i$ are constrained
to satisfy
 \be
 \sum_{i=1}^n \mu_i^{\,2}=1\,.
 \labell{direction}
 \ee
Hence not all of the $d\mu_i^2$ in the flat space metric \reef{flat}
are independent and one of these terms can be eliminated using this
constraint. However, we have left this replacement implicit for the
sake of keeping the metric simple. For completeness, we note that the
range of each of the coordinates is: $t\in(-\infty,\infty)$,
$r\in[0,\infty)$, $\mu_i\in[0,1]$ and $\phi_i\in[0,2\pi]$, where the
latter are periodically identified $\phi_i=\phi_i+2\pi$. We will adopt
polar coordinates analogous to those in eq.~\reef{flat} to present the
MP metrics for $d=2n+1$ below. In particular then, the black hole
geometry will approach the flat space metric \reef{flat}
asymptotically.

For an even number of dimensions, \ie $d=2n+2$ (with $n\ge1$), there
will be an extra unpaired spatial coordinate
 \be z=r\,\alpha \qquad {\rm with}\ \
\alpha\in [-1,1]\,.
 \labell{polar2}
 \ee
Hence the flat space metric becomes
 \be
ds^2= -dt^2 + dr^2 + r^2\,\sum_{i=1}^n \left(
d\mu_i^{\,2}+\mu_i^2\,d\phi_i^{\,2}\right)+ r^2\,d\alpha^2\,.
 \labell{flat2}
 \ee
while the constraint on the direction cosines becomes
 \be
 \sum_{i=1}^n \mu_i^{\,2} + \alpha^2=1\,.
 \labell{direction2}
 \ee
Eq.~\reef{flat2} exhibits the polar coordinates which we adopt below
for the MP metric with $d=2n+2$.

One outstanding feature of the polar coordinates in eqs.~\reef{flat}
and \reef{flat2} is that there are $n$ commuting Killing vectors in the
angular directions $\phi_i$. The corresponding rotations in each of the
orthogonal planes \reef{polar} match the $n$ generators of the Cartan
subalgebra of the rotation groups $SO(2n)$ or $SO(2n+1)$ for odd and
even $d$, respectively. This feature highlights the fact that in higher
dimensions we must think of angular momentum as an antisymmetric
two-tensor $J^{\mu\nu}$, \eg see eq.~\reef{spin}. In considering a
general rotating body, we may simplify this angular momentum tensor by
going to the center-of-mass frame, which eliminates the components with
a time index. Then a suitable rotation of the spatial coordinates
brings the remaining spatial components $J^{ij}$ into the standard form
 \be
J^{ij} = \pmatrix{\ 0&J_1&\ &\ &\ \cr
                 -J_1&0&\ &\ &\ \cr
                 \ &\ &\ 0&J_2&\ \cr
                 \ &\ &-J_2&0&\ \cr
                 \ &\ &\ &\ &\ddots\cr}\ .
 \labell{Jang}
 \ee
Here each of the $J_i$ denote the angular momentum associated with
motions in the corresponding plane. Note that for even $d$, the last
row and column of the above matrix vanishes. Therefore a general
angular momentum tensor is characterized by $n=\lfloor (d-1)/2\rfloor$
independent parameters $J_i$. Hence the general spinning black hole
metrics, which are considered below, will be specified by $n+1$
parameters: the mass $M$ and the $n$ commuting angular momenta $J^{y_i
x_i}$. In four dimensions, these parameters would completely fix the
black hole solution but, as we will see in section \ref{unstable} and
in subsequent chapters, these parameters alone will not fix a unique
black hole metric in higher dimensions.

\subsection{MP Black Hole Metrics} \labell{metrics}

As can be anticipated from eqs.~\reef{flat} and \reef{flat2}, the form
of the metrics differs slightly for odd and even dimensions. Hence let
us begin with the metric describing a spinning black hole in an even
number of spacetime dimensions, \ie $d=2n+2$ with $d\ge4$,
 \bea
ds^2 &=& - d{t}^2 + {\mu r\over\Pi\, F}\,
 \left(d{t}  + \sum_{i=1}^n a_i\, \mu_i^2 \,d{\phi}_i \right)^2
 + {\Pi\, F\over\Pi - \mu r} \,dr^2
  \non
 &&\qquad
+ \sum_{i=1}^n\,( r^2 + a_i^{\,2} ) \left(d\mu_i^{\,2} + \mu_i^2
\,{d{\phi}_i}^2 \right) + r^2\,d\alpha^2
 \labell{evenmet}
 \eea
where
 \bea
 F &=& 1-\sum_{i=1}^{n}\,{a_i^{\,2}\,\mu_i^{\,2} \over r^2 + a_i^{\,2}}\,
  \labell{FF}\\
 \Pi &=&\, \prod_{i=1}^{n}\, (r^2 + a_i^{\,2})\, . \labell{fpi}
 \eea
With $n=1$, we have $d=4$ and the above metric reduces to the well
known Kerr solution, discussed in Chapter 1.\footnote{To make the
connection more explicit, we would set $a_1=a$, $\mu_1=\sin\theta$ and
$\alpha=\cos\theta$.} For $d=2n+1$ with $d\ge5$, the metric becomes
 \bea
ds^2 &=& - d{t}^2 + {\mu r^2\over\Pi\, F}\,
 \left(d{t}  + \sum_{i=1}^n a_i\, \mu_i^2 \,d{\phi}_i \right)^2
 + {\Pi\, F\over\Pi - \mu r^2} \,dr^2
  \non
 &&\qquad
+ \sum_{i=1}^n\,( r^2 + a_i^{\,2} ) \left(d\mu_i^{\,2} + \mu_i^2
\,{d{\phi}_i}^2 \right)
 \labell{oddmet}
 \eea
with $F$ and $\Pi$ again given by eqs.~\reef{FF} and \reef{fpi}.
Examining the asymptotic structure of these metrics --- see
eq.~\reef{pert1} --- one finds that the $n$+1 free parameters, $\mu$
and $a_i$, determine the mass and angular momentum of the black hole
with
 \bea
M&=& {(d-2)\,\Omega_{d-2}\over16\pi G}\ \mu
 \labell{physical}\\
J^{y_i x_i} &=& {\Omega_{d-2}\over 8 \pi G}\, \mu\, a_i = {2\over
d-2}\, M\, a_i
 \nonumber
 \eea
where $\Omega_{d-2}$ is the area of an $S^{d-2}$ given in
eq.~\reef{solid}. Setting all of the spin parameters $a_i=0$, both
eqs.~\reef{evenmet} and \reef{oddmet} reduce to the $d$-dimensional
Schwarzschild metric \reef{sch}. Now also setting $\mu=0$ yields the
flat space metric in eqs.~\reef{flat} and \reef{flat2}, respectively.

With general spin parameters $a_i$, both metrics have $n$+1 commuting
Killing symmetries, corresponding to shifts in $t$ and $\phi_i$. These
symmetries are enhanced when some of the spin parameters coincide. In
particular, with $a_i=a$ for $i=1,\cdots,m$, the corresponding
rotational symmetry is enhanced from $U(1)^m$ to $U(m)$, where the
latter acts on the complex coordinates $z_i=\mu_i e^{i\phi_i}$ in the
associated subspace. A particularly interesting case is $d=2n+1$ with
all $n$ spin parameters equal. Then with the $U(n)$ symmetry, the
solution reduces to cohomogeneity-one, \ie it depends on a single
(radial) coordinate. Of course, if $k$ of the spin parameters vanish,
an $SO(2k)$ symmetry emerges in the corresponding subspace. When $d$ is
even, this enhanced rotational symmetry extends to $SO(2k+1)$ by
including the $z$ direction.

Of course, as with the Kerr metric, these geometries are only
stationary, rather than static, reflecting the rotation of the
corresponding black holes. In particular, the metric components $g_{t
\phi_i}$ are nonvanishing when $a_i\ne0$ and as a result, one finds
frame dragging in these higher dimensional spacetimes, just as was
described in Chapter 1 for four dimensions. 
We might also note that eqs.~\reef{evenmet} and \reef{oddmet} also
contain nonvanishing $g_{{\phi}_i{\phi}_k}$. Further, implicitly there
are also nonvanishing $g_{\mu_i \mu_k}$ (as well as $g_{\mu_i \al}$
with even $d$), which would appear explicitly if one of the direction
cosines were eliminated with eq.~\reef{direction} or \reef{direction2}.

%
%
\subsection{Singularities} \labell{sing}

Various components of the metrics, \reef{evenmet} and \reef{oddmet},
will diverge if either $\Pi\,F/r^{\ga}=0$ or $\Pi-\mu\, r^{\ga}=0$,
where $\ga=2$ and 1 for $d$ odd and even, respectively. The former
indicates a true curvature singularity while the latter corresponds to
an event horizon. To consider the former in more detail, one must
examine a list of separate cases, \ie odd or even $d$ and different
numbers of vanishing spin parameters. In most cases, one finds that
$\Pi\,F/r^{\ga}=0$ at $r=0$ and this entire surface is singular. There
are three exceptional cases which we consider in more detail below: a)
even $d$ and all $a_i\ne0$, b) odd $d$ and only one $a_i=0$, and c) odd
$d$ and all $a_i\ne0$. We should add that all of our comments with
regards to curvature singularities can be confirmed by directly
examining the behaviour of the curvatures. For example, we examine the
particular case of the $d=5$ MP metric in detail in Appendix B and our
results there explicitly match those discussed in (b) and (c) below.

 \vskip 0.5em
\noindent {\bf a) even $d$ and all $a_i\ne0$:} This case would include
the Kerr metric with $d=4$ and the results
are similar to those found there, as described in Chapter 1. 
First it is useful here to use the constraint \reef{direction2} to
re-express eq.~\reef{FF} as
 \be
F=\alpha^2+r^2\sum_{i=1}^n \frac{\mu_i^{\,2}}{r^2+a_i^2}\quad{\rm for\
even}\ d\,.
 \labell{feven}
 \ee
From this expression, we can see that in order for $\Pi\,F/r$ to vanish
we must have both $r=0$ and $\alpha=0$. Further intuition comes from
noting that it is most appropriate to think of the surfaces of constant
$r$ as describing ellipsoids of the form
 \be
\frac{z^2}{r^2}+\sum_{i=1}^n \frac{x_i^{\,2}+y_i^{\,2}}{r^2 +
a_i^{\,2}} =1\,.
 \labell{ellieven}
 \ee
For example, if we set $\mu=0$ in the black hole metric \reef{evenmet},
the resulting metric describes flat space foliated by these surfaces.
Hence as we approach $r=0$, these ($d$--2)-dimensional ellipsoids
collapse to a ($d$--2)-dimensional ball in the hyperplane $z=0$. Now
the direction cosine $\alpha=z/r$ acts as a radial coordinate in this
ball with $\alpha=1$ corresponding to the origin and $\alpha=0$ being
the boundary of the ball where the curvature diverges. Hence in higher
even dimensions, the ring-like singularity of the Kerr metric is
elevated to a singularity on a ($d$--3)-sphere. The ($d$--2)-ball at
$r=0$ acts as a two-sided aperture. Passing through the aperture to
negative values of $r$, we enter a new asymptotically flat space with
negative mass (and no
horizons). Further, as noted in Chapter 1 for the Kerr metric, 
this region also contains closed time-like curves. Passing through the
aperture a second time in the same direction, we reach a space
isometric to the original $r > 0$ region and for simplicity these two
regions are usually identified.

 \vskip 0.5em
\noindent {\bf b) odd $d$ and only one $a_i=0$:} For simplicity, let us
denote the vanishing spin parameter as $a_1$. We begin again by
rewriting eq.~\reef{FF}, this time using the constraint
\reef{direction}
 \be
F=\mu_1^2+r^2\sum_{i=2}^n \frac{\mu_i^{\,2}}{r^2+a_i^2}\quad{\rm for\
odd}\ d\ {\rm and}\ a_1=0\,.
 \labell{fodd}
 \ee
Hence in this case, for $\Pi\,F/r^2$ to vanish, we require both $r=0$
and $\mu_1=0$ --- note that $\Pi$ contributes a factor of $r^2$ here.
In this case, the appropriate geometric intuition comes from regarding
constant $r$ surfaces as ellipsoids of the form
 \be
\frac{x_1^{\,2}+y_1^{\,2}}{r^2}+\sum_{i=2}^n
\frac{x_i^{\,2}+y_i^{\,2}}{r^2 + a_i^{\,2}} =1\,.
 \labell{elliodd}
 \ee
Hence as we approach $r=0$, these ($d$--2)-dimensional ellipsoids
collapse to a ball in the hyperplane $x_1=0=y_1$. As above, $\mu_1$
acts as a radial coordinate in this ball with $\mu_1=0$ corresponding
to the boundary of the ball where the curvature diverges. However, a
key difference from the previous case is that here as $r\to0$, the
ellipsoids \reef{elliodd} become very narrow and collapse to a point in
the ($x_1,y_1$)-plane at $r=0$. Hence the ball at $r=0$ extends only in
$d$--3 dimensions. A careful examination of the geometry shows that
there is also a conical singularity in the ($x_1,y_1$)-plane for any
$\mu_1\ne0$.\footnote{Of course, this statement assumes that the mass
parameter $\mu$ is nonvanishing.} Hence the entire $r=0$ surface is in
fact singular here, although with a milder singularity than in the
generic cases.

 \vskip 0.5em
\noindent {\bf c) odd $d$ and all $a_i\ne0$:} If we apply the
constraint \reef{direction}, eq.~\reef{FF} becomes
 \be
F=r^2\sum_{i=1}^n \frac{\mu_i^{\,2}}{r^2+a_i^2}\quad{\rm for\ odd}\
d\,.
 \labell{fodd2}
 \ee
In this case, we observe that $\Pi$ approaches a finite constant at
$r=0$ and eq.~\reef{direction} does not allow all of the $\mu_i$ can
vanish simultaneously. Hence, $\Pi F/r^2$ remains finite at $r=0$ and
so there is no curvature singularity here. However, the metric
\reef{evenmet} remains problematic at this location since one finds
that $g_{rr}\propto r^2$ as $r\to0$. However, this is only a coordinate
singularity which is avoided by choosing a new radial coordinate
$\rho=r^2$. Now in passing to negative values of $\rho$, the function
$\Pi F/r^2(\rho)$ eventually vanishes and a curvature singularity
arises at $\rho=-a_s^2$, where $a_s$ is the absolute value of the spin
parameter(s) with the smallest magnitude. If more than one spin
parameter has the value $\pm a_s$, the entire surface $\rho=-a_s^2$ is
singular. If only one spin parameter, say $a_1$, has the value $\pm
a_s$, the singularity at $\rho=-a_s^2$ only appears at $\mu_1=0$. In
this case, if $a_{s'}$ is the absolute value of the next smallest spin
parameter, the geometry extends smoothly to values of
$-a_{s'}^2\le\rho\le -a_s^2$ in certain directions. However, the
curvature singularity extends throughout this range of $\rho$ since $F$
can vanish for certain angular directions. Hence ultimately all
trajectories moving towards smaller values of $\rho$ end on a
singularity in this region.


%
%
\subsection{Horizons} \labell{housefire}

In considering the event horizons for these metrics, we must again
consider separately the cases where the spacetime dimension is even or
odd. Let us start with $d=2n+2$, which includes the Kerr metric for
$d=4$. The event horizons arise where $g^{rr}$ vanishes and so from
eq.~\reef{evenmet}, we require
 \be
\Pi-\mu r =0\,.
 \labell{evenhor}
 \ee
Thus the horizons correspond to the roots of a polynomial, which is
order $d-2$ in $r$. Unfortunately, apart from $d=4$ or 6, there will be
no general analytic solutions (in terms of radical expressions) for the
position of the horizon. Hence a complete set of necessary and
sufficient conditions for the existence of a horizon is unavailable for
higher $d$. However, we can still make some general observations.

First of all if it exists the horizon must have the topology of
$S^{d-2}$ since it is a surface of constant $r$. Further to avoid a
naked singularity, we require the mass (\ie $\mu$) to be positive. The
latter can be deduced with two observations: first, the singularity
appears at $r=0$ and second, the function $\Pi$ is everywhere positive
(or zero) --- recall the definition in eq.~\reef{fpi}. Hence for
eq.~\reef{evenhor} to have a root at positive $r$, we must have
$\mu>0$. With a closer examination of the polynomial in
eq.~\reef{evenhor}, we see that, in fact, it is large and positive for
large $|r|$ and has a single minimum. Hence we conclude that there are
only three possible scenarios: two, one or zero horizons. Hence in this
regard, the higher dimensional metrics \reef{evenmet} are the same as
the familiar Kerr metric in $d=4$. However, an interesting difference
arises if one (or more) of the spin parameters vanishes. Recall that
$Pi$ is monotonically increasing and grows as $r^{2n}$ at large $r$.
However, in this case, $\Pi$ vanishes at $r=0$ and grows as $r^{2m}$
for small $r$, with $m$ vanishing spin parameters. Hence the right-hand
side of eq.~\reef{evenhor} is negative for small $r$ while it still
becomes large and positive for large $r$. Hence there must always be
one nondegenerate root at positive $r$, corresponding to a single
horizon. This result holds irrespective of how large the remaining spin
parameters are and hence the event horizon appears even when the
angular momentum grows arbitrarily large, as long as there is no
rotation in at least on of the orthogonal planes. These solutions with
very large angular momenta have been dubbed `ultra-spinning' black
holes in \cite{one}. As we will see in section \ref{unstable}, the
latter have further interesting consequences.

For $d=2n+1$, the location of the horizon in eq.~\reef{oddmet} is
determined by
 \be
 \Pi - \mu r^2=0\, .\labell{evenhorz}
 \ee
It is more useful to present this expression using the new radial
coordinate $\rho=r^2$ introduced in the previous discussion of
singularities. In terms of $\rho$, eq.~\reef{evenhorz} becomes
 \be
\prod_{i=1}^{n}(\rho+{a_i}^2)-\mu \rho=0\, .\labell{evenhorz2}
 \ee
Hence we are looking for the roots of a polynomial of order $n$ and so
analytic solutions only exist for $n=2$, 3 and 4, \ie d=5, 7, 9 ---
these are given in Appendix B for $d=5$. Of course, the horizon has the
topology of $S^{d-2}$ since it is a surface of constant
$\rho$.\footnote{Implicitly we are assuming $\rho>0$ here. See the
additional discussion below of the case where all of the spin
parameters are nonvanishing.} Finding a root with $\rho>0$ again
requires positive $\mu$. In fact, a positive root requires
 \be
 \mu>\sum_i\prod_{j\ne i}{a_j}^2\,,
 \labell{necess}
 \ee
which ensures that the coefficient of the linear term is negative in
eq.~\reef{evenhorz2}. This constraint is necessary but not sufficient
for the absence of a naked singularity. Provided that $\mu$ is
sufficiently large, we will again only find one or two horizons with
positive $\rho$, just as in the case of even $d$. Note that for odd
$d$, a single vanishing spin parameter is insufficient to guarantee the
existence of a horizon, since the constraint \reef{necess} remains
nontrivial. However if two or more of the spin parameters vanish,
eq.~\reef{evenhorz2} has one positive root, as well as a root at
$\rho=0$. Further in this particular case, we can have regular
ultra-spinning solutions where the event horizon appears even when the
remaining spin parameters become arbitrarily large.

Recall that the singularity structure distinguished the case of odd $d$
and all $a_i\ne0$. In particular, in this case, the surface $\rho=0$ is
nonsingular and the geometry extends to negative values of $\rho$. To
avoid naked singularities here, we only need that the outermost
horizon, \ie the largest root of eq.~\reef{evenhorz2}, appears for
$\rho>-a_s^2$ where the singularity appears.\footnote{As in the
previous section, to discuss this case, we adopt the notation that
$a_s$ and $a_{s'}$ are the magnitudes of the smallest and second
smallest spin parameters, respectively.} Now with positive $\mu$, the
only possibility is that the horizon appears at positive $\rho$
provided $\mu$ is sufficiently large, as described above. On the other
hand, we have $\Pi(\rho=-a_s^2)=0$ and hence for any negative $\mu$, a
root appears in eq.~\reef{evenhorz2} in the range $-a_s^2<\rho<0$.
Below, we will see that these negative mass solutions are even more
pathological since they contain causality violating regions extending
beyond the horizon. To close this discussion, we recall that when only
one spin parameter has the minimal value, the geometry extends further
to the range $-a_{s'}^2<\rho<-a_s^2$. In this case, for small positive
$\mu$, one finds two roots or one degenerate root in this new range.
However, these surfaces intersect the singular surface and so the
latter is not entirely concealed by these horizons. Further if horizons
occur in the range $-a_{s'}^2<\rho<-a_s^2$, one may show no other
horizons appear for positive $\rho$. Therefore these spacetimes contain
naked singularities.

\subsection{Ergosurfaces and Causality Violation} \labell{ergoz}

Turning now to ergosurfaces, we must determine the surfaces where
$g_{tt}$ vanishes. From the metrics in eqs.~\reef{evenmet} and
\reef{oddmet}, the latter correspond to the roots of
 \bea
F\Pi - \mu r &=&0\,, {\rm\qquad even\ }d\non
F\Pi -\mu r^2&=&0\,, {\rm\qquad odd\ }d \labell{ergo}
 \eea
for $r>0$. These surfaces still have the topology of $S^{d-2}$ but, of
course, the factor $F$ introduces a more complicated directional
dependence than appears for the horizons. As above, while there is no
analytic solution for these equations, one is still able to deduce the
general properties of the surfaces. In particular, one such surface
always appears outside of the outer horizon and another may appear
inside the inner horizon, if the latter exists. As can be seen from
eq.~\reef{ergo}, the ergosurface will touch the horizons where $F=1$.
If $m$ spin parameters vanish when $d$ is even, then the latter
corresponds to the (2$m$)-dimensional sphere described by
$1=\alpha^2+\sum_{k=1}^m \mu_k^2$, where the sum runs over the $m$
indices for which $a_k=0$. Hence if no spin parameters vanish, the two
surfaces only touch at the two points on the horizon where $\alpha=\pm
1$, as found for the four-dimensional Kerr metric. Similarly if $m$
spin parameters vanish when $d$ is odd, the ergosurface and horizon
will touch along the $S^{2m-1}$ described by $1=\sum_{k=1}^m \mu_k^2$.
In particular, the two surfaces will not coincide anywhere if all of
the spin parameters are nonvanishing in the case of odd $d$. Further in
this case, one finds that with positive $\mu$, there will be an
ergosurface outside of the outer horizon but no such surface inside the
inner horizon. On the other hand, if $\mu$ is negative, no ergosurfaces
exist at all.

As described in Chapter 1, 
the outer ergosurface marks the boundary within which particles cannot
remain at rest with respect to infinity. Further, the spinning black
holes in higher dimensions can be mined with Penrose processes, just as
in four dimensions. Another analogy with $d=4$ arises in the scattering
waves propagating in these geometries, which produces superradiance for
the MP solutions as in the Kerr metric.

We close this section by turning to the question of causality
violation. For many of the black holes under consideration, we need
only consider $r>0$ and in this domain, the angular coordinates are
perfectly well-behaved. The exceptional cases requiring additional
consideration correspond to the black holes where all of the $a_i\ne0$.
First for even $d$, $r$ can be extended to negative values in the
second asymptotic region. In this region, the metric components
$g_{\phi_i\phi_i}$ can become negative leading to closed time-like
loops, as occurs in the Kerr metric. For odd $d$ and all $a_i\ne0$, the
geometry extends beyond $r=0$ to negative values of $\rho=r^2$. In this
case for a each angle $\phi_i$, eq.~\reef{oddmet} gives
 \be
g_{\phi_i\phi_i}=(\rho+{a_i}^2)\left( 1+{\mu {a_i}^2\over\Pi} \right)
 \labell{ctc}
 \ee
in the plane $\mu_i=1$. The above expression will become negative if
the second factor has a zero, \ie for radii inside that where
$\Pi+\mu{a_i}^2=0$. Now recall that with $\mu<0$, the horizon arises at
the root of eq.~\reef{evenhorz2} which lies between $\rho=-a_s^2$ and
0. Hence the more important observation is that for any angle $\phi_i$
for which the corresponding spin parameter satisfies $a_i^2>a_s^2$, the
above metric component will be negative for some values of $\rho$
outside of the horizon (since $\Pi$ is a monotonically increasing
function). That is, the negative mass solutions typically contain
causality violating regions extending beyond the horizon --- the only
exception would be the case when all of the spin parameters are
precisely equal. For completeness, we also note that in this case with
$\mu>0$ and a single $a_i$ taking the value $\pm a_s$, there is the
possibility that eq.~\reef{ctc} may vanish for $a_i=a_s$ in the range
$- a_{s'}^2< \rho< -a_s^2$.

%
%
\subsection{Maximal Analytic Extension} \labell{max}

In examining the maximal analytic extension of the solutions
\reef{evenmet} and \reef{oddmet}, one can use the usual techniques
developed to study four-dimensional black holes and the results are
essentially the same as for $d=4$. In particular, one finds two
separate extensions of the spacetime at each horizon, \ie an infalling
coordinate patch which extends the geometry across the future horizon
and an outgoing patch which smoothly traverses the past horizon. In the
following, our discussion will focus on the case of even $d$ and the
extension of eq.~\reef{evenmet}. However, with the obvious changes, the
same discussion is easily adapted to the case of odd $d$, as we briefly
examine near the end of this section.

Towards the construction of the maximal analytic extension of these
spacetime geometries, it is straightforward to construct Eddington-like
coordinates
 \bea
 dt &= &d{t_\pm} \mp
 {\mu r\over \Pi - \mu r} \,dr\,,
 \labell{edding}\\
d{\phi}_i &=& d\phi_{\pm,i} \pm {\Pi\over\Pi - \mu r}\, {a_i \,dr\over
r^2 + {a_i}^2}\nonumber\,.
 \eea
With these new coordinates, the metric \reef{evenmet} becomes
 \bea
 ds^2 &=& -dt_\pm^2 + dr^2  + \sum_{i=1}^n
 ( r^2 + {a_i}^2 ) \left(d\mu_i^{\,2} + \mu_i^{\,2}
\,d\phi_{\pm,i}^2 \right) + r^2 \,d\alpha^2 \non
 &&\ \ \pm 2\sum_{i=1}^n a_i\,\mu_i^{\,2}  \,d\phi_{\pm,i} \,dr + {\mu
r\over\Pi\, F} \left(dt_\pm \pm dr + \sum_{i=1}^n a_i\, \mu_i^{\,2}
\,d\phi_{\pm,i} \right)^2\, .\labell{eddy2}
 \eea
Hence the metric is well-behaved in either coordinate system at the
horizons, \ie $\Pi-\mu r=0$. Of course, various metric components are
still singular at $\Pi\,F/r=0$ since the latter corresponds to a true
curvature singularity. As can be seen from eq.~\reef{eddy2}, each of
these coordinate systems are adapted to a particular family of radial
geodesics following the null vectors
 \be
 k_\pm^\mu{\partial\ \over\partial x^\mu}={\partial\ \over\partial t_\pm}
\mp{\partial\ \over\partial r}\,.\labell{geo1}
 \ee
That is, the `+' and `--' coordinates are well-behaved along infalling
and outgoing geodesics, respectively, which cross the horizons. Hence
$t_+$ remains finite on the future horizon, where $r\to r_\mt{H}$ and
$t\to+\infty$, while $t_-$ remains finite on the past horizon, where
$r\to r_\mt{H}$ and $t\to-\infty$.

The above Eddington-like coordinates \reef{edding} indicate that the
structure of the horizons is essentially the same as that found in four
dimensions. In particular, let us consider the case where
eq.~\reef{evenhor} has two distinct roots at positive $r$ --- recall
this requires that all of the spin parameters are nonvanishing. Hence
we have an outer event horizon at $r=r_{\mt{H}}$ and an inner Cauchy
horizon at $r=r_{\mt{C}}$ ($<r_{\mt{H}}$). The corresponding Penrose
diagram is shown in figure \ref{penrose}. A typical Eddington
coordinate patch covers three regions in this diagram: the
asymptotically flat exterior region where $r>r_{\mt{H}}$; the central
region between the inner and outer horizons where
$r_{\mt{C}}<r<r_{\mt{H}}$; and the inner region where $r<r_{\mt{C}}$
which contains a time-like ``ring" singularity and which can be
extended to an asymptotically flat region (with $r<0$). If we consider
the regions covered by the infalling coordinates $\lbrace
t_+,\phi_{+,i}\rbrace$, then each of these three regions can be
separately extended by transforming to the outgoing coordinates,
$\lbrace t_-,\phi_{-,i}\rbrace$. Hence the maximally extended spacetime
becomes a tower in which the basic geometry illustrated in figure
\ref{penrose} is repeated an infinite number of times. We might note
that, as illustrated in the figure, the horizons at $r=\rh$ and
$r_\mt{C}$ have the characteristic `X' structure of a bifurcate Killing
horizon. Here the various branches of the horizon are connected at the
bifurcation surface at the center of the X, which corresponds to a
fixed point of the associated Killing vector. Strictly speaking to
demonstrate that the regions of the various overlapping Eddington
patches are in fact smoothly connected at the bifurcation surface, one
should find Kruskal-like coordinates, which are simultaneously
well-behaved across both the future and past horizons (as well as the
bifurcation surface). While this is certainly possible, the
construction of these coordinates is a more involved exercise and we
refer the interested reader to \cite{merry} for a discussion of this
point.
\begin{figure}[!ht]
  \centering  {\includegraphics[width=0.5\textwidth]{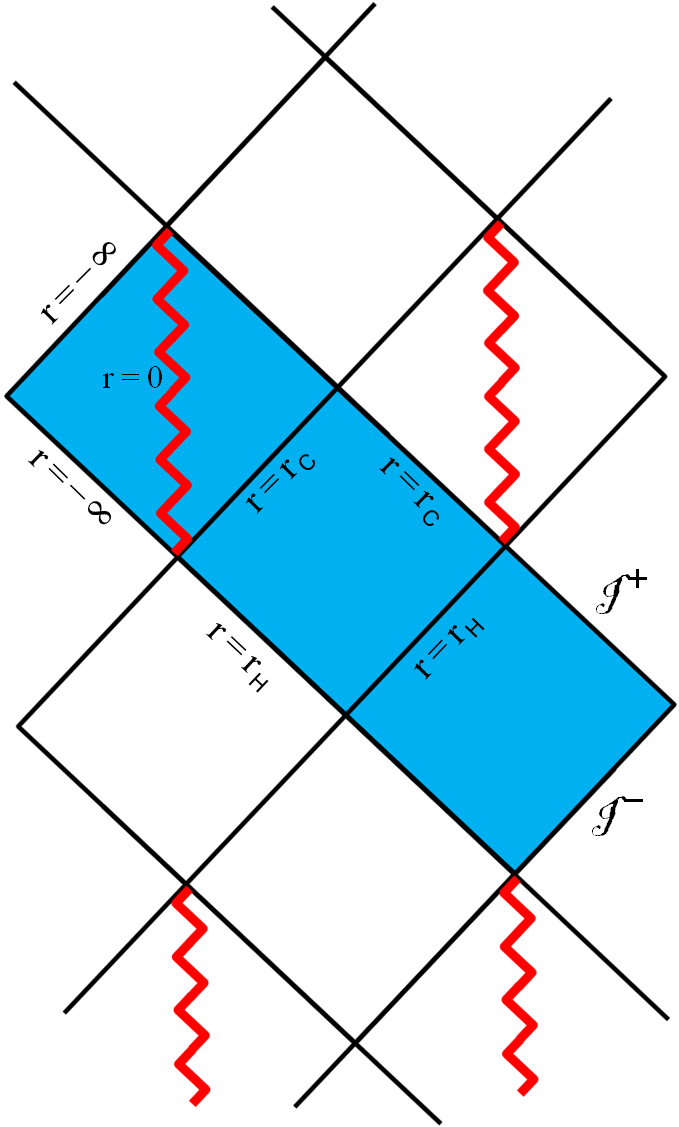}}
\caption{Penrose diagram for spinning black hole with two horizons for
even $d$. The shaded regions indicate a single coordinate patch covered
by infalling Eddington coordinates. \labell{penrose}  }
\end{figure}

As noted above the inner horizon at $r=r_{\mt{C}}$ is a Cauchy horizon,
representing the boundary for the unique evolution of initial data on
some space-like surface stretched across the Einstein-Rosen bridge
joining two asymptotically flat regions. Now we expect that these
Cauchy horizons should be unstable since the same simple arguments,
which indicate such a surface is unstable in the four-dimensional Kerr
metric, can be applied equally well here in higher dimensions. However,
it must be said that this issue has not been studied in the same detail
as in four dimensions and so an accurate description of the resulting
singularity remains lacking for higher dimensions.

Above, we considered the spinning black holes \reef{evenmet} (with all
of the $a_i\ne0$) in the regime where there were two distinct horizons.
Now if the mass of this solution is fixed and some of the spin
parameters are increased, eventually the two horizons will coalesce
producing an extremal black hole. In this case, the individual
Eddington coordinate patches cover the exterior region and the inner
region, and connecting these patches results in the maximal extension
illustrated in figure \ref{penrose2}(a). In this case, the near-horizon
analysis of \cite{gary} can also be extended to higher dimensions to
find that the throat region of the extremal black hole corresponds to
an analog of the geometry $AdS_2\times S^n$ \cite{texas}. If any of the
spin parameters are further increased then the horizon disappears and
one is left with a naked singularity, as shown in figure
\ref{penrose2}(b). Hence the extended black hole geometries described
here and above provide a direct analogue in higher dimensions of the
four-dimensional story for the Kerr solution, described in Chapter 1.
\begin{figure}[!ht]
  \centering  {\includegraphics[width=0.8\textwidth]{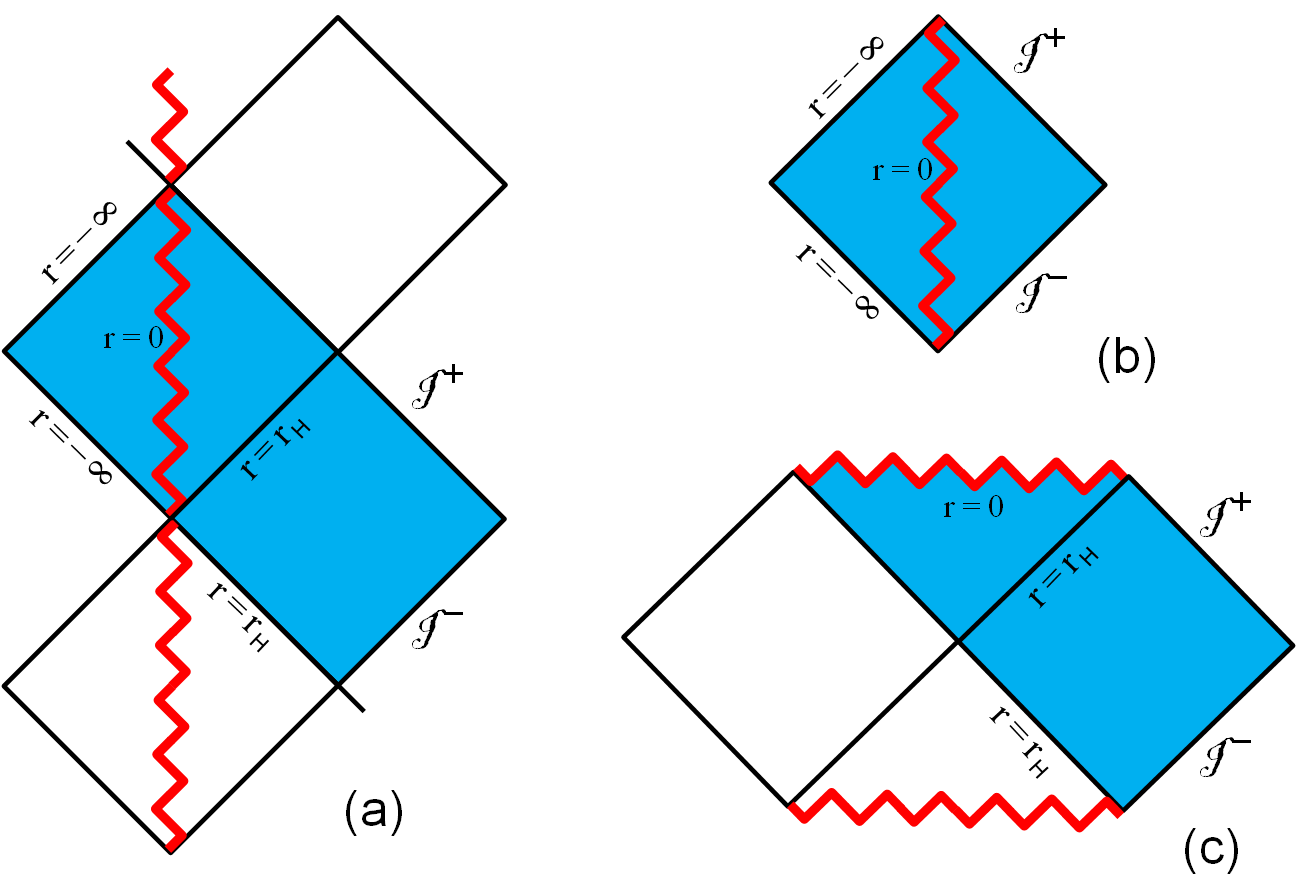}}
\caption{Further Penrose diagrams for even $d$: a) an extremal spinning black hole with single
degenerate horizon, b) an over-rotating solution without horizon, and c) a spinning
black hole with one or more $a_i=0$. As before,
 the shaded regions indicate a single coordinate patch covered
by infalling Eddington coordinates.
\labell{penrose2}  }
\end{figure}

Another possibility, which we have not yet considered for even $d$, is
when one or more of the spin parameters vanish. In this case, there is
a single horizon but that it corresponds to a simple zero in
eq.~\reef{evenhor}. There will be a second root but it occurs at the
singularity at $r=0$. One finds that this singular surface is
space-like and so the Penrose diagram is similar to that of the
Schwarzschild solution. In particular, there is no infinite tower of
connected regions here but rather the singularities form space-like
boundaries for the future and past interior regions. Here an analogy
might be drawn with the $d=4$ Kerr metric in the limit that $a\to0$
where $r_\mt{C}\to0$ and the geometry reduces to the Schwarzschild
solution. However, in higher dimensions, there will in general still be
other nonvanishing spin parameters but the structure of the spacetime
remains unchanged irrespective of how large the remaining $a_i$ become.
Hence, as noted above, with one $a_i=0$ (and $d$ is even), we can
construct ultra-spinning black holes carrying an arbitrary amount of
angular momentum.

The above discussion was restricted to even $d$ but there are no
essential differences for the case of odd $d$. Of course as mentioned
in section \ref{sing}, with all of the spin parameters nonvanishing,
the surface $r=0$ is nonsingular and the geometry extends to negative
values of $\rho=r^2$. Further one finds a time-like singularity in the
latter domain but there is no connection to a second asymptotically
flat region. Another difference is that the cases where the Penrose
diagram takes a Schwarzschild form includes either two or more $a_i$=0
and $\mu>0$ or one $a_i$=0 and $\mu>\sum_i\prod_{j\ne i}{a_j}^2$. The
same structure also arises when all $a_i\ne 0$ and $\mu<0$ but, as
described above, these spacetimes are pathological since they contain
causality violating regions outside of the horizon.

To close this section, let us make a few supplementary comments. First,
we note that the metrics in eq.~\reef{eddy2} actually have the
so-called Kerr-Schild form
 \be
 g_{\mu\nu}=\eta_{\mu\nu}\ +\ h\,(k_\pm)_\mu\,(k_\pm)_\nu
 \labell{ks}
 \ee
where $h=\mu r/\Pi F$. Of course, a further coordinate transformation
would be required to introduce Cartesian coordinates so that the flat
space line-element takes the conventional form. Here I might note that
one of the remarkable features of the four-dimensional Kerr metric is
that it can be written in this particular form \cite{kerr}. Ultimately,
it was the fact that the MP metrics can also be written in the
Kerr-Schild form that allowed us to derive eqs.~\reef{evenmet} and
\reef{oddmet}.

It is also interesting to examine the null vectors \reef{geo1} in the
original coordinate system given in eq.~\reef{evenmet}:
 \be
k_\pm^\mu{\partial\ \over\partial x^\mu}={\Pi\over\Pi-\mu r}\left(
{\partial\ \over\partial t}-\sum_{i=1}^n\omega_i{
\partial\ \over\partial\phi_i}\right)\mp{\partial\ \over\partial r}
 \labell{vec2}
 \ee
where $\omega_i={a_i\over r^2+{a_i}^2}$. From these expressions, we see
that upon approaching the horizon,
 \be
k_\pm^\mu{\partial\ \over\partial x^\mu}\propto {\partial\
\over\partial t}-\sum_{i=1}^n \Omega_i{
\partial\ \over\partial\phi_i}\,,
 \ee
with $\Omega_i={a_i\over \rh^2+{a_i}^2}$. That is, $k_-^\mu$ becomes
the generator of the future horizon at $r=r_\mt{H}$ in the infalling
Eddington coordinate patch described by $\lbrace
t_+,\phi_{+,i}\rbrace$. Similarly with infalling Eddington coordinates,
$k_+^\mu$ matches the generator of the past horizon at $r=r_\mt{H}$. A
final comment is that these two vector fields given in eq.~\reef{geo1}
or \reef{vec2} correspond to the principal null vectors that appear in
the algebraic classification, discussed in Chapter 9.

\subsection{Hidden Symmetries and Geodesics} \labell{hidden}

In the four-dimensional Kerr metric, particle motion is easily studied
because the geodesics are completely soluble by quadratures. That is,
there are four constants of motion, which allow us to write the
complete solution for geodesic motion in terms of a set of indefinite
integrals. At first sight, this is a rather remarkable property since
the Killing symmetries and the fixed norm of the four-velocity only
provide three such constants. The fourth constant is more subtle and
relies on the existence of a Killing-Yano tensor in this particular
background \cite{pen2} -- see below. The existence of this tensor is
also responsible for the separability of the wave equation for spin-0,
-1/2, -1 and -2 fields in this background. Recent work uncovered a rich
structure of analogous relationships in higher dimensions, \eg
\cite{frolov,don,commute,sepp}. In particular, the required hidden
symmetries were found for the Myers-Perry metrics \cite{frolov}, from
which one can infer the integrability of geodesic motion in these
backgrounds \cite{don}.

Central to this discussion is the existence of a rank-two closed
conformal Killing-Yano tensor (CCKY) $h_{\mu\nu}$ which is a two-form
satisfying
 \be
\nabla_{(\mu}h_{\nu)\rho}={1\over d-1}\left(g_{\mu\nu} \,\nabla_\sigma
h^\sigma{}_\rho -\nabla_\sigma
h^\sigma{}_{\!(\mu}\,g_{\nu)\rho}\right)\,.
 \labell{yano}
 \ee
As this two-form is closed, it also satisfies $dh=0$ and so at least
locally there exists a one-form potential $b$ such that $h=db$. In the
case of the MP metrics, \reef{evenmet} and \reef{oddmet}, the CCKY
tensor can be explicitly written as
 \bea
h & =& \sum^n_{i=1}\, a_i\,\mu_i\,d\mu_i\wedge \left( a_i \,dt + (r^2 +
a_i^{\,2} )\, d\phi_i\right)\non
 &&\qquad +\ r\,dr\wedge\left( dt + \sum^n_{i=1} a_i\,\mu_i^2 \,
d\phi_i\right)\, . \labell{explicit}
 \eea

Following the standard construction in four dimensions, one constructs
a second-rank Killing tensor \cite{pen2}
 \be
K_{(\mu\nu)}=-h_\mu{}^\sigma\,h_{\nu\sigma}+\frac{1}{2}g_{\mu\nu}h_{\rho\sigma}
h^{\rho\sigma}
 \labell{toot}
 \ee
which then satisfies the identity
 \be
\nabla_{\!(\mu}K_{\nu\rho)}=0\,.
 \labell{kill3}
 \ee
It follows then that along a geodesic described by the $d$-velocity
$u^\mu$, the following is a constant of the motion:
$K_{\mu\nu}\,u^\mu\, u^\nu$. In higher dimensions, the latter is only
the first in a series of new conserved quantities. We will not describe
the construction here but one finds the following tower of second-rank
Killing tensors \cite{frolov,don}
 \bea
 K^{(\ell\,)\,\mu}{}_\nu&=&
{(2\ell)!\over(2^\ell \,\ell!)^2}\left( \delta^\mu{}_\nu\,
h^{[\mu_1\nu_1}\cdots h^{\mu_\ell\nu_\ell]}h_{[\mu_1\nu_1}\cdots
h_{\mu_\ell\nu_\ell]}
 \right.\non
 &&\left.\qquad\qquad-2\ell\,h^{\mu[\nu_1}\cdots h^{\mu_\ell\nu_\ell]}
h_{\nu[\nu_1}\cdots h_{\mu_\ell\nu_\ell]}\right)\,. \labell{new}
 \eea
Note that comparing this expression to eq.~\reef{toot}, we see
$K^{(1)}_{\mu\nu}=K_{\mu\nu}$. Now using eq.~\reef{yano} for the CCKY
tensor, it follows that all of these tensors satisfy the identity
\reef{kill3} and hence each provides a constant of the motion along a
geodesic: $c_\ell=K^{(\ell)}_{\mu\nu}\,u^\mu\, u^\nu$.

From the above expression \reef{new}, it appears that this construction
extends to $\ell=1,\cdots,n+1$ for $d=2n+2$. However, one finds that
for $\ell=n+1$ that the right-hand side vanishes as an identity. On the
other hand, one naturally extends this series to $\ell=0$ with
$K^{(0)\,\mu}{}_\nu=\delta^\mu{}_\nu$, in which case
$c_0=K^{(0)}_{\mu\nu}\,u^\mu\, u^\nu=g_{\mu\nu}\,u^\mu\, u^\nu$ is
simply the norm of the $d$-velocity. Hence the Killing tensors then
provide $n+1$ constants of motion. An essential feature of this
construction is that these constants are in fact all independent. The
latter statement is related to the fact that the CCKY tensor contains
$n+1$ independent `eigenvalues' for even $d$, when it is put in the
standard form analogous to eq.~\reef{Jang}. Of course, the Killing
symmetries (time translations and the $n$ rotations in each $\phi_i$)
provide a further $n+1$ constants of the motion. Hence in total, there
are $d=2n+2$ constants which allow us to solve for the geodesics in
quadratures.

For $d=2n+1$, there is a similar counting of the constants of motion.
In this case, the Killing tensors provide $n+1$ independent constants
 $c_\ell$ with $\ell=0,1,\cdots,n$. Further the Killing
symmetries provide $n+1$ independent constants. At this point, it may
seem that we have too many integration constants but, in the case of
odd $d$, it turns out that $c_n$ is reducible. That is,
$c_{(n)}=(\xi^\nu\,g_{\mu\nu}\,u^\mu)^2$ where $\xi^\nu$ is a Killing
vector \cite{don}. This result is related to describing eq.~\reef{new}
as the contraction of a CCKY tensor of rank $d-2\ell$ (dual to the
wedge product of $\ell$ $h$'s). Hence for $\ell=n$, the latter is a
one-form for which the analog of eq.~\reef{yano} reduces to Killing's
equation. Hence this tensor is in fact simply a linear combination of
the Killing vectors. Consequently, the total number of independent
constants is precisely $d=2n+1$ and the geodesic motion is again
completely integrable \cite{don}.

We comment that it has also been shown that the Killing(-Yano) tensors
also lead to the separability of the Klein-Gordon and Dirac equations,
as well as the Hamilton-Jacobi equations in these backgrounds, \eg
\cite{sepp}. While we do not have room to describe these results in
detail here, a key element in this analysis is to construct `symmetry
operators' which commute with the appropriate wave operator. For
example, in the case of the Klein-Gordon equation \cite{commute}, we
can start with simple operators constructed for each of the Killing
coordinates, \ie $i\,\partial_t$ and $i\,\partial_{\phi_i}$, each of
which commute with $\nabla^2-m^2$. Various components of the separated
solution of $(\nabla^2-m^2)\psi=0$ can then be identified as
eigenfunctions of these operators, \eg $e^{i\omega t}$ and
$e^{im\phi_i}$. Now the Killing tensors provide an additional set of
symmetry operators: $\hat{K}^{(\ell)}=\nabla_\mu
(K^{(\ell)\,\mu\nu}\nabla_\nu)$, which also satisfy
$[\nabla^2-m^2,\hat{K}^{(\ell)}]$. Again, various separated components
of the desired solutions can then be written as eigenfunctions of these
new operators. It remains an open question as to whether a similar set
of symmetry operators can be constructed for the field equations of a
Maxwell field or linearized gravitons and whether separability extends
to these equations. We might note that some progress in analyzing
linearized metric perturbations has been made for the particular case
of odd $d$ and all $a_i$ equal \cite{idea}.

\subsection{Black Hole Thermodynamics} \labell{thermo}

As already commented in chapter 1, the basic framework of black hole
thermodynamics extends from four to higher dimensions in a
straightforward way. We might add that implicitly this relies on the
fact that our discussion of higher dimensional black holes is
restricted to solutions of Einstein's equations. There have also been
interesting extensions of black hole thermodynamics to include both
higher curvature actions and higher dimensions \cite{wald}. In any
event, we will keep our comments here brief --- see also comments in
the following section.

The zeroth law, namely, that the surface gravity or temperature (\ie
$T=\kappa/2\pi$) is constant across any stationary event horizon, is
essential if the corresponding black holes are to behave like a thermal
bath. This result is easily established if the horizon is a bifurcate
Killing horizon, which is certainly the case here, following the
discussion of section \ref{max}. As noted there, the horizon generator
is given by
 \be
 \chi^\mu\,\partial_\mu=\partial_t-\sum_{i=1}^n\,\Omega_i\,\partial_{\phi_i}\,.
 \labell{genera}
 \ee
Recall that $\Omega_i={a_i\over \rh^2+{a_i}^2}$. Hence using
$\chi^\sigma\nabla_\sigma \chi^\mu=\kappa \chi^\mu$ to evaluate the
surface gravity, one finds
 \be
\kappa=\cases{ {\partial_r\Pi-\mu\over2\mu r}\Big\vert_{r=r_\mt{H}}&\
for even $d$\,,\cr {\partial_r\Pi-2\mu r\over 2\mu
r^2}\Big\vert_{r=r_\mt{H}}&\ for odd $d$\,.\cr}
 \labell{casek}
 \ee
While these are somewhat formal expressions, they clearly illustrate
that $\kappa$ is constant across the entire horizon.

Of course, the first law takes precisely the same form as in four
dimensions:
 \be
\delta M={\kappa\over8\pi G}\, \delta{\cal A}+\sum_{i=1}^n\Omega_i\,
\delta J_i
 \labell{firstl}
 \ee
which leads to the interpretation of the area of (a cross-section of)
the horizon $\cal A$ as the entropy of the black hole with the
celebrated formula: $S={\cal A}/4G$. (Of course, in an $d$-dimensional
spacetime, this area $\cal A$ actually has the dimensions of length to
the power $d-2$.) The Killing symmetries of the MP metrics also allow
us to construct a useful related relation, known as the integrated
Smarr formula \cite{smarr},
 \be
 \frac{d-3}{d-2}\,M=\sum_{i=1}^n\,\Omega_i\,J_i+\frac{\kappa}{8\pi G}
\,{\cal A}\, .
 \labell{smarr}
 \ee

Following \cite{smarr}, the irreducible mass of the black hole may be
identified from the first law. This is the mass associated with the
area of the horizon, \ie one integrates the area term in
eq.~\reef{firstl},
 \bea
M_{ir}&=&{1\over 8\pi G}\int_0^{\cal A}\kappa({\cal A}^\prime, J_i=0)\,
d{\cal A}^\prime\non
 &=&{d-2\over16\pi G}\,\Omega_{d-2}^{\,1/(d-2)}\,{\cal
A}^{d-3\over d-2}\non
 &=&{d-2\over d-3}\,{\kappa\, {\cal A}\over8\pi G}
 \eea
Hence $M-M_{ir}$ is the mass or energy connected to the rotation of the
black hole and we expect that it may be removed through Penrose
processes. In four dimensions, this can be explicitly verified because
the geodesics in the Kerr metric are completely soluble by quadratures.
Given the recent developments described in section \reef{hidden}, it
would be interesting to extend this analysis to higher dimensions.

To close this section, we note that the second law (\ie
$\delta\cA\ge0$) is also easily extended to higher dimensions,
following the discussion in Chapter 1. 
One proof of the latter relies on the matter falling across the horizon
satisfying the null energy condition and also on cosmic censorship
\cite{hellis}. While the former still seems a reasonable assumption in
higher dimensions, the latter may appear more dubious given the
recent results discussed in Chapter 3. 
However, the second law may also be proved by using the null energy
condition and by demanding that the null generators of the horizon are
complete \cite{hellis}. In fact, the latter is consistent with our
current understanding of the final state of the Gregory-Laflamme
instability and hence it seems that the second law remains to have a
firm foundation in higher dimensions.

\subsection{Instabilities} \labell{unstable}

While there is strong evidence for the stability of Kerr black holes in
four dimensions, in fact, the opposite is true for spinning black holes
in higher dimensions. That is, we believe that in higher dimensions,
various instabilities arise for MP black holes when the angular
momentum becomes large. In fact, it has been argued that these
instabilities are related to the appearance of a rich fauna of new
black holes in higher dimensions \cite{phase,threeb}.

A precise understanding of instabilities would require an analysis of
the linearized perturbations of the MP metrics, \reef{evenmet} and
\reef{oddmet}. While this is possible in four dimensions, as noted in
section \ref{hidden}, limited progress has been made in higher
dimensions. However, insight into the situation in higher dimensions
comes from making connections with the Gregory-Laflamme instability of
black branes  --- see Chapter 2. As described below, this approach led
to the conjecture that ultra-spinning black holes should be unstable
for $d\ge6$ \cite{one} and numerical evidence of this conjecture was
recently found \cite{three,lost,four}. An interesting consequence is
that it seems that general relativity in higher dimensions imposes a
dynamical `Kerr bound' on the spin  of the form $J^{d-3}\lesssim G
M^{d-2}$ in $d$ dimensions.

To illustrate this point, let us consider the spinning black hole
solutions with a single nonvanishing spin parameter. With this
restriction, for either odd or even $d$, the metric reduces to
\begin{eqnarray}
ds^2&=& -dt^2 + \frac{\mu}{r^{d-5}\rho^2}\left( dt+a\sin^2\theta
\,d\varphi\right)^2 +{\Sigma\over\Delta}dr^2
 \labell{mphole}\\
&&\qquad+\Sigma\,d\theta^2
+(r^2+a^2)\sin^2\theta\, d\varphi^2 + r^2\cos^2\theta\, d\Omega^2_{d-4}\,,
 \nonumber
\end{eqnarray}
where
\beq \Sigma=r^2+a^2\cos^2\theta\qquad{\rm and}\qquad
\Delta=r^2+a^2-\frac{\mu}{r^{d-5}}\,. \eeq
Here we have set $a_1=a$ and $\mu_1=\sin\theta$ (as well as
$a_{i>1}=0$). Now the event horizon is determined as the largest root
$r_\mt{H}$ of $\Delta(r)=0$. That is,
\beq \rh^2+a^2 -\frac{\mu}{\rh^{d-5}}=0\,.\labell{horz}\eeq
In examining this equation, it is not hard to see that for $d=4$ or 5,
there is an extremal limit (\ie an upper bound on $a$) beyond which no
horizon exists. However, as our discussion in section \ref{housefire}
indicated, the more interesting case is $d\ge6$. For the latter, we may
note that the term $r^2$ makes the left-hand side of eq.~\reef{horz}
large and positive as $r\to\infty$. On the other hand, the term
$-\mu/r^{d-5}$ makes $\Delta(r)$ negative for small $r$ and hence there
must be a (single) positive root independent of the value of $a$. That
is, we have the possibility of ultra-spinning solutions, for which a
regular event horizon remains even when the angular momentum (per unit
mass) grows arbitrarily large.

Let us examine the geometry of the horizon of eq.~\reef{mphole} in this
ultra-spinning regime. In the limit of very large $a$ and fixed $\mu$,
the solution of eq.~\reef{horz} is approximately given by
\beq \rh\simeq \left(\frac{\mu}{a^2}\right)^{1/(d-5)}\ll a\,.
\labell{rplus}\eeq
Hence we observe that $\rh$ is shrinking as $a$ grows (and $\mu$ is
kept fixed). However, $\rh$ is simply some coordinate expression and
one must instead examine the horizon in a covariant way to uncover the
true geometry. Various approaches may be taken here, all with the same
simple result. If we characterize the size of the horizon along and
orthogonal to the plane of rotation as $\ell_\parallel$ and
$\ell_\perp$, respectively, then
 \be
\ell_\parallel \sim a \quad{\rm and}\quad \ell_\perp \sim r_{\mt H}\,.
 \labell{sizes}
 \ee
That is, the horizon of these rapidly rotating black holes spreads out
in the plane of rotation while contracting in the transverse
directions, taking a `pancake' shape in this plane. Considering the
area of the horizon, we find
\beq \cA=\Omega_{d-2} \rh^{d-4}(\rh^2+a^2) \simeq \Omega_{d-2}
\rh^{d-4}a^2 \simeq \Omega_{d-2}\left(
\frac{\mu^{d-4}}{a^2}\right)^{1/(d-5)}\,. \labell{atot}\eeq
Note that the area decreases as $a$ grows because the contraction in
the transverse directions overcomes the spreading in the plane of
rotation. We emphasize that this result \reef{atot} only applies for
$d\ge6$. The horizon area also decreases with increasing $a$ in $d=4$
or 5, but it is only for larger $d$ that we can consider the
ultra-spinning regime with $a\to\infty$, in which case the area shrinks
to zero size.

Hence from the perspective of an observer near the axis of rotation and
near the horizon (\ie near $\theta\sim0$ and $r\sim r_\mt{H}$), the
horizon geometry appears similar to that of a black
membrane,\footnote{This statement can be made mathematically precise in
the limit $a\to\infty$ \cite{one}.} \ie it has roughly the geometry
$R^2\times S^{d-4}$. However, as we saw in
Chapter 2, 
Gregory and Laflamme found that a black membrane would be classically
unstable when the size in the brane directions is larger than that of
the transverse sphere \cite{GL}. Hence it is natural to expect that the
ultra-spinning MP solutions are unstable in the limit $a\to\infty$ but
also that the instability actually sets in at some finite value of $a$
\cite{one}.

The transition between the horizon behaving similar to the Kerr black
hole and behaving like a black membrane is easily seen using black hole
thermodynamics. One simple quantity to consider is the black hole
temperature of the metric \reef{mphole}. Beginning from zero spin, $T$
decreases as $a$ grows, just like in the familiar case of the Kerr
black hole. In $d=4$ and $d=5$ the temperature continues to decrease
reaching zero at extremality, however, in $d\geq 6$ there is no
extremal limit. So instead, $T$ reaches a minimum and then starts
growing again, as expected for a black membrane. The minimum, where
this behavior changes, can be determined exactly  \cite{one}
 \beq
\left.\frac{a^2}{\rh^2}\right|_{crit} = \frac{d-3}{d-5}
 \qquad {\rm or} \qquad
\left.\frac{a^{d-3}}\mu\right|_{crit}=\frac{d-3}{2(d-4)} \,\left(
\frac{d-3}{d-5}
 \right)^{\frac{d-5}2}\,.
 \labell{transit}
 \eeq
Following \cite{three}, we can use this critical ratio \reef{transit}
to define the boundary of the ultra-spinning regime. That is,
ultra-spinning solutions are defined to be those for which the ratio
$a^{d-3}/\mu$ exceeds the critical value given in eq.~\reef{transit}.
Explicitly evaluating eq.~\reef{transit} for the latter ratio, we finds
some of these critical values to be
 \beq
\left.\frac{a^{d-3}}{\mu}\right|_{crit}= 1.30,\ 1.33,\ 1.34,\ 1.35\quad
{\rm for}\ d=6,\ 7,\ 8,\ 9,\ {\rm respectively}\,. 
 \labell{ratio}\eeq
We note that these critical values seem to be only weakly dependent on
$d$. Further, these results would seem to indicate that the
membrane-like behaviour, and hence the instability, arises for
relatively small values of the spin parameter $a$.

A further connection to black hole thermodynamics appears because it is
expected that the classical Gregory-Laflamme instabilities should be
connected to thermodynamic instabilities of the corresponding black
branes \cite{GM}. More precisely, it was conjectured that the
appearance of a negative `specific heat' for the black brane is
connected to the appearance of this classical instability. Applying
this reasoning in the present context would suggest that the rotating
black hole should become unstable at some point after
$\partial^2S/\partial J^2>0$ \cite{phase}, \ie after the point of
inflection marked `x' in figure \ref{phasediag}. Given the expression
for the area \reef{atot}, one finds that this point corresponds
precisely to that identified above from the behaviour of the
temperature. That is, the critical point `x' where
$\partial^2S/\partial J^2=0$ is given precisely by eq.~\reef{transit}.
\begin{figure}[!ht]
  \centering  {\includegraphics[width=0.5\textwidth]{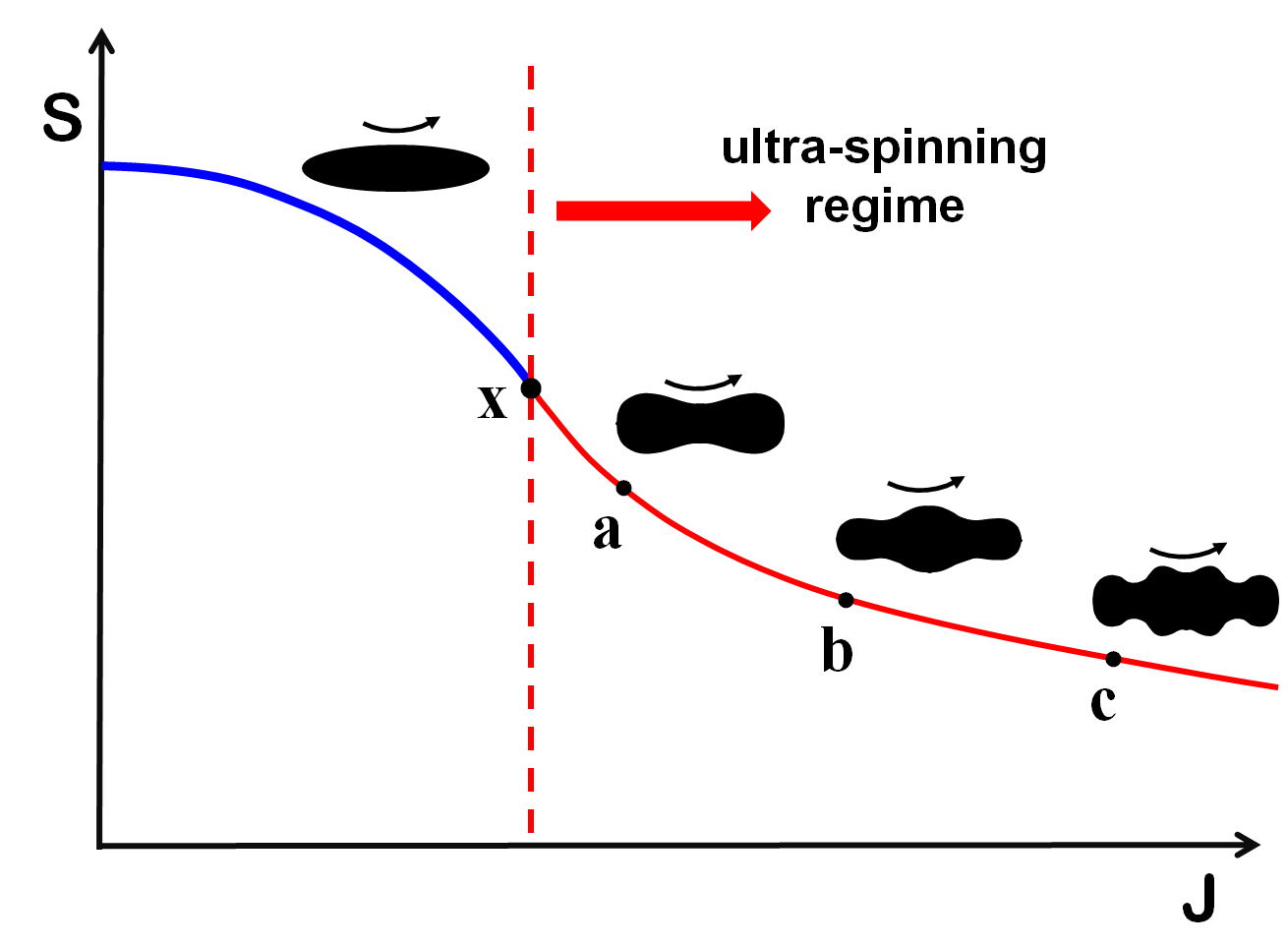}}
\caption{Phase diagram of entropy vs. angular momentum, at fixed mass, for MP black holes spinning
in a single plane for $d \ge 6$. The point `x' indicates where $\partial^2S/\partial J^2 = 0$.
The subsequent points (a,b,c, $\ldots$) correspond to the threshold of axisymmetric instabilities
which introduce increasing numbers of ripples in the horizon. It is further conjectured that
a new class of black holes with rippled horizons branches off from each of these points \cite{phase}.
\labell{phasediag}}
\end{figure}

While resolving these issues analytically remains intractable at
present, there has been remarkable progress coming from numerical
investigations in recent years \cite{three}. If one considers the
instability just at threshold, \ie precisely at the critical value of
$a$, then the corresponding frequency is precisely zero and the
unstable mode becomes a time-independent zero-mode. In \cite{three},
with a particular ansatz for such zero-modes, the authors were able to
numerically locate the corresponding critical values of $a$ for the
singly spinning MP black holes \reef{mphole} in $d=6$ to 11. In fact,
they found such a mode precisely where $\partial^2S/\partial J^2 = 0$.
However, the interpretation of this stationary mode is more subtle.
Rather than corresponding to an instability, this perturbation simply
corresponds to shifting the solution to a nearby MP black hole with a
slightly larger spin. However, a small distance further into the
ultra-spinning regime, they were also found a new zero-mode which
`pinches' the horizon at the axis of rotation, as illustrated for the
point `a' in figure \ref{phasediag}. It is believed that this zero-mode
does correspond to the onset of a true instability for higher values of
the angular momentum $J$. Further, this was only the first of a
hierarchy of zero-modes which introduced an increasing number of
pinches or ripples in the event horizon along the $\theta$ direction.
While these numerical searches only identified the stationary modes (by
design), this provides strong evidence for a hierarchy of
Gregory-Laflamme instabilities in the ultra-spinning regime.

These zero-modes also provide evidence for a new class of stationary
rotating black holes with spherical horizons but with a rippled profile
in the polar angle $\theta$. The existence of these solutions was also
conjectured in \cite{one,phase}. According to the phase diagram
suggested in \cite{phase}, there would be a new branch of solutions
beginning at the point `a' and in moving along this branch, the pinch
in the horizon at the axis of rotation would grow larger and larger.
The conjecture is that this branch connects to yet another phase where
the pinch produces a puncture in the horizon and the new phase would
consist of spinning
black rings, analogous to those discussed in Chapter 6 
except the horizon topology would be $S^1\times S^{d-3}$. Similarly, it
is conjectured that the branch starting from the point `b' would
connect the spinning MP black holes to higher dimensional versions of
the `black saturn' found in \cite{saturn} for five dimensions. Hence
the new spinning black holes with rippled spherical horizons appear
only to be a precursor to a rich fauna of new solutions with complex
horizon topologies in higher dimensions

Implicitly, the latter analysis was only considering modes which
respect all of the rotational symmetries present in the original metric
\reef{mphole}, \ie $U(1)\times SO(d-3)$. However, this restriction was
only imposed to simplify the analysis. A priori, there is no reason why
all of the unstable modes should respect these symmetries. In fact,
recent numerical studies suggest that non-symmetric modes play an
important role in these instabilities. In \cite{four}, full numerical
simulations were carried out to describe evolution of rapidly spinning
MP black holes in higher dimensions --- again with a single
nonvanishing spin parameter as in eq.~\reef{mphole}. In all of the
cases studied, it was found that the solutions were unstable against
non-axisymmetric perturbations, with an initial profile proportional to
$\sin(2\phi)$. The critical value where this `bar-mode' instability set
in was found to be:
 \beq
\left.\frac{a^{d-3}}{\mu}\right|_{bar}= 0.76,\ 0.41,\ 0.28,\ 0.27\quad
{\rm for}\ d=5,\ 6,\ 7,\ 8,\ {\rm respectively}\,.
 \labell{ratio2}\eeq
We should note that these values are considerably smaller than those
identified above, in eq.~\reef{ratio}. Notably, these numerical
simulations were able to find an instability of the $d=5$ MP black
hole, where the previous discussion was unable to identify any
instabilities. Further, following the nonlinear evolution of the
unstable perturbation, the simulations \cite{four} found that the
deformed black holes spontaneously emit gravitational waves causing
them to spin down and settle again to a stable MP black hole with a
spin parameter smaller than the critical value in eq.~\reef{ratio2}. An
open question is to determine when such `bar-mode' instabilities arise
for MP black holes rotating in more than one plane. As an aside, let us
note here that in $d = 5$ with both spin parameters equal, it was shown
analytically that no instabilities appear whatsoever \cite{blossom}.

In the preceding discussion, we have only considered MP black holes
rotating in a single plane. However, this was only done to simplify the
presentation and because this case was the focus of the numerical
studies in \cite{three,four}. As discussed in section \ref{housefire},
ultra-spinning black hole solutions can also arise with several of
nonvanishing spin parameters growing large, as long as one (or two) of
the spin parameters vanish in even (or odd) $d$. It is natural to
expect that the ultra-spinning regime also extends to the regime where
several $a_i$ grow large while the remainder stay small. Guided by this
intuition, it is straightforward to extend the original discussion of
the Gregory-Laflamme-like instabilities to the case where several spin
parameters, say $m$, grow without bound while the remainder stay finite
(or vanish) \cite{one}. The limiting metric describes a (rotating)
black 2$m$-brane, where the horizon topology is $R^{2m}\times
S^{d-2-2m}$. However, a Gregory-Laflamme-like instability is again
expected to appear for these branes when the characteristic size in the
planes with large spins is somewhat larger than the characteristic size
in the transverse directions. In general, with many independent spins,
the thermodynamic analysis mentioned above extends studying of the
Hessian $\partial^2S/\partial J_i \partial J_j$ for negative
eigenvalues \cite{three}. This expression provides a more refined
definition of ultra-spinning black holes. In particular, following the
discussion with a single nonvanishing $J_i$, we define the boundary of
the ultra-spinning regime as the boundary where this Hessian first
acquires a zero eigenvalue.

Further insights into ultra-spinning instabilities have been found for
one other example \cite{threeb,blossom,threec}, namely, odd $d=2n+1$
with all of the $n$ spin parameters equal. As noted, in section
\ref{metrics}, the rotational symmetry of these geometries is enhanced
to $U(n)$ and it can be shown that the metric involves a fibration over
the complex projective space $CP^n$ \cite{idea}. Further the metric
perturbations of these spacetimes can be decomposed as harmonics on
this $CP^n$ and their analysis reduces to the study of an ordinary
differential equation for the radial profile. Of course, in these
metrics with all $a_i\ne0$, there is an extremal limit and so it is not
immediately obvious that one can reach an ultra-spinning regime or that
any instabilities should appear. In fact, analysis of the above Hessian
reveals an ultra-spinning regime for any odd $d\ge7$.
Ref.~\cite{threeb} explicitly identified unstable modes for $d=9$ and
supplementary work \cite{threec} later found unstable modes appeared
very close to the extremal limit for $d=7,$ 9, 11 and 13.
Ref.~\cite{blossom} was able to show that no instabilities arise for
$d=5$. Hence these results suggest that instabilities will arise in
these cohomogeneity-one black hole spacetimes for any odd $d\ge7$.
Recently these instabilities of the cohomogeneity-one black holes were
connected to those of the singly spinning black holes with the
numerical work of ref.~\cite{lost}. They showed that the ultra-spinning
instabilities in these two sectors are continuously connected by
examining perturbations of MP black holes with all but one of the spin
parameters being equal. While their explicit calculations were made for
$d=7$, similar results are expected for higher odd $d$ as well.

To close, we observe that the construction of the threshold zero-modes
in $d=9$ suggest that there should be a new family of spinning black
hole solutions characterized by 70 independent parameters
\cite{threeb}!! Generically, these solutions would have only two
Killing symmetries, \ie time translations and one $U(1)$ rotation
symmetry. Hence here again, the ultra-spinning instabilities open the
window on a exciting panorama of new black hole solutions in higher
dimensions.

\section*{Acknowledgements}

Research at Perimeter Institute is supported by the Government of
Canada through Industry Canada and by the Province of Ontario through
the Ministry of Research \& Innovation. The author also acknowledges
support from an NSERC Discovery grant and funding from the Canadian
Institute for Advanced Research. I would also like to thank the Aspen
Center for Physics for hospitality while preparing this paper. I would
also like to thank Roberto Emparan, Gary Horowitz, David Kubiznak and
Jorge Santos for their comments on this manuscript.

\section*{Appendix A: Mass and Angular Momentum}

This appendix will consider the definition of the mass and angular
momentum of an isolated gravitating system in $d$ dimensions. Our
approach is to simply generalize the standard asymptotic analysis of
four-dimensional solutions of Einstein's equations \cite{ADM} to higher
dimensions. In particular, the mass and angular momentum of any
isolated gravitating system (\eg a black hole) may be defined by
comparison with a system which is both weakly gravitating and
non-relativistic. The result then provides the \dd-dimensional
generalization of the ADM mass and angular momentum \cite{ADM}.

So let us begin with the \dd-dimensional Einstein equations
 \be
R_{\mu \nu} - {1\over 2} g_{\mu \nu} R = 8 \pi G \, T_{\mu \nu} \,,
 \labell{eom}
 \ee
where we have included the stress-energy tensor for some matter fields,
as it will be useful in the following discussion.\footnote{In the
following, Greek indices run over all values $\mu, \nu = 0,1,\ldots
d-1$, while Latin indices only run over spatial values $ i, j =
1,2,\ldots d-1$.} Now we wish to consider solutions of these equations
when the gravitating system is both weakly gravitating and
non-relativistic. First, with a weakly gravitating system, the metric
is everywhere only slightly perturbed from its flat space form:
 \be g_{\mu \nu} = \eta_{\mu \nu} + h_{\mu \nu}\,, \labell{alm}\ee
with $\vert h_{\mu \nu} \vert \ll 1$. Next, if the the system is
non-relativistic, any time derivatives of fields will be much smaller
than their spatial derivatives. Of course, this also implies that
components of the stress energy tensor may be ordered
 \be
\vert T_{00}\vert \gg \vert T_{0i}\vert \gg\vert T_{ij}\vert\,.
 \labell{stress}
 \ee
These inequalities indicate that the dominant source of the
gravitational field is the energy density while the momentum density
provides the next most important source.

The solutions are most conveniently examined in the harmonic gauge
 \be
\partial_\mu\left(h^{\mu\nu}-{1\over 2}\eta^{\mu\nu} {h^\alpha}_\alpha\right)
=0\,.
 \labell{harm}
 \ee
With this choice, to leading order, the Einstein equations \reef{eom}
can be written as
 \bea
 \nabla^2 h_{\mu \nu} &=& - 16 \pi G \left( T_{\mu \nu} - {1\over d-2}
 \eta_{\mu \nu} T^\alpha{}_\alpha\right)\non
    &=& - 16 \pi G\ \widetilde{T}_{\mu \nu}
 \labell{linear}
 \eea
where $\nabla^2$ is the ordinary Laplacian in flat \dd-dimensional
space, \ie we have dropped the time derivatives of the metric
perturbation.  Note that $T^\al{}_\al \approx - T_{00}$ for
non-relativistic sources. Eq.~\reef{linear} is now readily solved with
 \be
h_{\mu \nu}(x^i) = {16 \pi G\over (N-2) \Omega_{d-2}} \int
{\widetilde{T}_{\mu \nu}(y^i)\over \vert \vec{x} - \vec{y}
\vert{}^{d-3}}\ d^{d-1}\! y
 \labell{result}
 \ee
where the integral extends only over the $(d-1)$ spatial directions.
Recall that $\Omega_{d-2}$ denotes the area of a unit (\dd--2)-sphere,
as given in eq.~\reef{solid}. Now evaluating eq.~\reef{result} in the
asymptotic region far from any sources, we have $r = \vert \vec{x}
\vert \gg \vert \vec{y} \vert$ and so we may expand the result as
 \be
  h_{\mu \nu}(x^i) = {16 \pi G\over (d-3) \Omega_{d-2}}\,
{1\over r^{d-3}} \int \widetilde{T}_{\mu \nu}\,d^{d-1}\! y  + {16 \pi
G\over \Omega_{d-2}}\,{x^k\over r^{d-1}} \int y^k\, \widetilde{T}_{\mu
\nu}\,d^{d-1}\! y
 + \cdots
  \labell{expand}
  \ee

To simplify our results, we consider the system in its rest frame,
which implies
 \be\int T_{0i} \,d^{d-1}\! x = 0\,, \labell{stop}\ee
and we choose the origin to sit at the center of mass, which fixes
 \be \int x^k\, T_{00} \,d^{d-1}\! x = 0,. \labell{zero}\ee
Now the total mass and angular momentum are defined as
 \bea
M&=& \int T_{00} \,d^{d-1}\! x \,,\labell{mass}\\
J^{\mu \nu}&=&\int(x^\mu T^{\nu 0}-x^\nu T^{\mu 0})\,d^{d-1}\!x\,.
 \labell{spin}
 \eea
One further simplification comes from the conservation of
stress-energy, which reduces to $\partial_kT^{k\mu } = 0 $ in the
present case of interest and from which we can infer
 \be
\int x^\ell\, T^{k \mu} \,d^{d-1}x  =
 - \int x^k \,T^{\ell\, \mu} \,d^{d-1}\!x\, . \labell{jj7}
 \ee
Now this result, along with eqs.~\reef{stop} and \reef{zero}, allows us
to simplify the angular momentum to
 \bea
 J^{0k} &=& 0 \non
 J^{kl} &=& 2 \int x^k\, T^{\ell\,0} \,d^{d-1}\! x\labell{spin2}
\, .
 \eea

Applying these results to the expansion in eq.~\reef{expand}, we find
that to leading order far from the system
 \bea
  h_{00} &\approx& {16 \pi G\over(d-2) \Omega_{d-2}}\,
 {M\over r^{d-3}}\,, \non
  h_{ij} &\approx& {16 \pi G\over (d-2)(d-3) \Omega_{d-2}}\,
 {M\over r^{d-3}}\, \delta_{ij}\,, \labell{pert1}\\
  h_{0i} &\approx& - {8 \pi G\over \Omega_{d-2}}\, {x^k \over r^{d-1}} J^{ki}
\,. \nonumber
 \eea
While these results were derived for a system which is both weakly
gravitating and non-relativistic, the asymptotic behaviour of the
metric will be the same for any isolated gravitating system. In
particular then, we use these expressions to identify the mass and
angular momentum of the black hole solutions discussed in the main
text.

\section*{Appendix B: A Case Study of d=5}

For $d=5$ dimensions, we can write the metric \reef{evenmet} as
 \bea
ds^2 &=& - d{t}^2 + {\mu \over\Sigma}\,
 \left(d{t}  + a\, \sin^2\theta \,d{\phi}_1 + b\, \cos^2\theta \,d{\phi}_2\right)^2
 + {r^2\Sigma\over\Pi - \mu r^2} \,dr^2
  \non
 &&\qquad
+ \Sigma\, d\theta^2 + (r^2+a^2)\sin^2\theta \,d{\phi}_1^{\,2} +
(r^2+b^2)\cos^2\theta \,d{\phi}_2^{\,2}
 \labell{met5}
 \eea
where
 \bea
 \Sigma&=& r^2 + a^{2}\cos^2\theta+ b^{2}\sin^2\theta\,,
  \labell{sigma5}\\
 \Pi &=&\,  (r^2 + a^{2})\,(r^2 + b^{2})\, . \labell{fpi5}
 \eea
Comparing our notation here to that in the main text, we  have set
$a_1=a$, $a_2=b$, $\mu_1=\sin\theta$ and $\mu_2=\cos\theta$.

 \vskip 0.5em
\noindent{\bf Singularities:} Now with some computer assistance, one
can easily calculate the Kretchman invariant
 \be
R_{\mu\nu\rho\sigma}R^{\mu\nu\rho\sigma}=\frac{24 \mu^2}{\Sigma^{\,6}}
(4r^2-3\Sigma)\,(4r^2-\Sigma)\,.
 \labell{kretch}
 \ee
At $r=0$, this expression yields
 \be
\left. R_{\mu\nu\rho\sigma}R^{\mu\nu\rho\sigma}\right|_{r=0}=\frac{72
\mu^2}{(a^{2}\cos^2\theta+ b^{2}\sin^2\theta)^4} \,.
 \labell{kretch0}
 \ee
Hence if $b=0$ above, we see there is a divergence as $\theta\to\pi/2$,
as described in case (b) in section \ref{sing}. Further with $b=0$, if
we examine $(r,\phi_2)$ part of the metric near $r=0$ but away from
$\theta=\pi/2$, we find
 \be
 ds^2\simeq \frac{\cos^2\theta}{1-\frac{\mu}{a^2}}\left(dr^2 +\left(1-
 \frac{\mu}{a^2}\right)\,r^2\,d\phi_2^{\,2}\right)
 +\cdots\,.
 \labell{cone5}
 \ee
Hence we see that there is an angular deficit of
$\Delta\phi_2=2\pi\mu/a^2$ on this axis.

On the other hand with both $a$ and $b$ nonvanishing, curvature
invariant in eq.~\reef{kretch0} remains finite. In this case, we
introduce the radial coordinate $\rho=r^2$ and assuming $0<a^2\le b^2$,
we find
 \be
\left.
R_{\mu\nu\rho\sigma}R^{\mu\nu\rho\sigma}\right|_{\rho=-a^2}=\frac{24
\mu^2(4a^2+3(b^2-a^2)\sin^2\theta)(4a^2+(b^2-a^2)\sin^2\theta)}{(b^2-a^2)^6\,\sin^{12}\theta}
\,.
 \labell{kretch1}
 \ee
Hence in accord with the discussion of case (c) in section \ref{sing},
the surface $\rho=-a^2$ is entirely singular if $b^2=a^2$. However, if
$b^2\ne a^2$, the singularity in eq.~\reef{kretch1} only appears at
$\theta=0$. Thus in this case, we can extend the geometry into the
region $-b^2\le\rho\le-a^2$. However, one finds that for any value of
$\rho$ in this domain, there are singularities at
 \be
\sin^2\theta=\frac{|\rho|-a^2}{\,b^2-a^2}
 \labell{clover}
 \ee
where $\Sigma=0$.

 \vskip 0.5em
\noindent{\bf Horizons:} With $d=5$, eq.~\reef{evenhorz} for the
horizon becomes a quadratic equation in $r^2$ and the roots are given
by the relatively simple expressions
 \bea
2{\rh}^2&=&\mu-{a}^2-{b}^2+\sqrt{(\mu-{a}^2-{b}^2)^2-4{a}^2
{b}^2}\,,\labell{hori5}\\
2{r_\mt{C}}^2&=&\mu-{a}^2-{b}^2-\sqrt{(\mu-{a}^2-{b}^2)^2-4{a}^2
{b}^2}\,.\nonumber
 \eea
Therefore the existence of a horizon requires
 \bea
\mu&\ge&{a}^2+{b}^2+2\,\vert\, a\,b\,\vert\non
M^3&\ge&{27\,\pi\over32\,G}({J_1}^2+{J_2}^2+2\vert\, J_1J_2\vert) \, .
 \labell{hori5b}
 \eea
The definitions of the mass and angular momentum given by
eq.~\reef{physical} have been inserted to yield the second equation and
we have defined $J_1\equiv J^{y_1x_1}$ and $J_2\equiv J^{y_2x_2}$.
Hence there are no ultra-spinning black holes in $d=5$. Rather, if the
angular momentum exceeds the above condition \reef{hori5b}, the
solution contains a naked `ring' singularity without any event horizon.

 \vskip 0.5em
\noindent{\bf Ergosurfaces:} The equation for the ergosurface reduces
to $\Sigma-\mu=0$ or
 \be
r_\mt{E}^2(\theta)=\mu - a^2\,\cos^2\theta-b^2\,\sin\theta^2\,.
 \labell{poit}
 \ee
When both $a$ and $b$ are nonvanishing, it is not hard to show that
$r_\mt{E}^2>\rh^2$, \ie the ergosurface nowhere touches the horizon.

 \vskip 0.5em
\noindent{\bf Cohomogeneity-One:} It is also interesting to observe the
simplifications that arise when $b=a$. First note that in this case, we
have
 \be
 \Sigma= r^2 + a^{2}\,,\quad{\rm and}
 \quad
 \Pi =  (r^2 + a^{2})^2 . \labell{fpi5aa}
 \ee
Further then, we see that the angular components in the second line of
eq.~\reef{met5} now combine to give $(r^2+a^2)\,d\Omega^2_3$, \ie the
round metric on a three-sphere. Hence this portion of the metric is
symmetric under $SO(4)\simeq SU(2)\times SU(2)$. However, this symmetry
does not survive for the full metric because there are other angular
contributions in the first line of eq.~\reef{met5}. However these terms
can be written in terms of the potential
 \be
A=i(\bar{z}_1\,dz_1+ \bar{z}_2\,dz_2)=
\sin^2\theta\,d\phi_1+\cos^2\theta \,d\phi_2\,, \labell{pottt}
 \ee
where $z_1=\sin\theta\,e^{i\phi_1}$ and $z_2=\cos\theta\,e^{i\phi_2}$.
Writing $A$ in terms of these complex coordinates makes clear that the
surviving symmetry is $U(1)\times SU(2)=U(2)$, as discussed in section
\ref{metrics}. The metric \reef{met5} with $b=a$ is called
cohomogeneity-one because after imposing this $U(2)$ symmetry, the
metric components are entirely functions of the single (radial)
coordinate $r$.

This enhanced symmetry also leads to a simplicity in other aspects of
the geometry. For example, the Kretchman invariant \reef{kretch} is now
only a function of $r$,
 \be
R_{\mu\nu\rho\sigma}R^{\mu\nu\rho\sigma}=\frac{24 \mu^2}{(r^2+a^2)^{6}}
(r^2-3a^2)\,(3r^2-a^2)\
\simeq_{\!\!\!\!\!\!\!_{{}_{r\to0}}}\frac{72\mu^2}{a^{8}}\,.
 \labell{kretchaa1}
 \ee
Hence the singularity at $\rho=-a^2$ in eq.~\reef{kretch1} simplifies
to
 \be
\left.
R_{\mu\nu\rho\sigma}R^{\mu\nu\rho\sigma}\right|_{\rho=-a^2+\varepsilon^2}=\frac{384
\mu^2\,a^2}{\varepsilon^{12}} \,,
 \labell{kretch1aaaa}
 \ee
where we are assuming that $\varepsilon\ll a$. We might also note that
for these black holes, the location of ergosurface \reef{poit} reduces
to $r_\mt{E}^2=\mu - a^2$ and so the latter is now also independent of
$\theta$. Given this simple result, it is also a straightforward
exercise to write
 \be
r_\mt{E}^2-r_\mt{H}^2=\frac\mu2\left(1-\sqrt{1-{4a^2}/\mu}\right)>0\,,
 \labell{diffeh}
 \ee
confirming that the ergosurface does not touch the horizon at any point
in these cohomogeneity-one black hole spacetimes.

%

\end{document}